\documentclass[conference]{IEEEtran}
\IEEEoverridecommandlockouts
\usepackage{graphicx}
\usepackage[prologue,usenames,table,dvipsnames,x11names]{xcolor}
\usepackage{tabularx, makecell, booktabs}
\usepackage{cite}
\usepackage{amsmath,amssymb,amsfonts,mathtools}
\usepackage{framed}
\usepackage[most]{tcolorbox}
\usepackage{algorithmic}
\usepackage{textcomp}
\usepackage{xspace}
\usepackage[abbreviations]{foreign}
\usepackage[inline]{enumitem}
\usepackage[breaklinks,colorlinks=true,linkcolor=black,anchorcolor=black,citecolor=black,filecolor=black,menucolor=black,runcolor=black,urlcolor=black]{hyperref}
\usepackage{pifont}
\usepackage{numprint}
\usepackage{caption}
\usepackage{url}
\usepackage{breakurl}
\usepackage[varqu]{zi4}
\usepackage{orcidlink}
\usepackage[export]{adjustbox}
\usepackage{eso-pic}
\usepackage[most]{tcolorbox}

\definecolor[named]{ACMDarkBlue}{cmyk}{1,0.58,0,0.21}

\usepackage{hyperxmp}
\hypersetup{
    pdfauthor={Jämes Ménétrey, Marcelo Pasin, Pascal Felber, Valerio Schiavoni},
    pdftitle={WaTZ: A Trusted WebAssembly Runtime Environment with Remote Attestation for TrustZone},
    linkcolor=ACMDarkBlue,
    citecolor=ACMDarkBlue,
    urlcolor=ACMDarkBlue,
}

\def\bmLatenciesTimeNativeTaTimeMean{10}

\def\bmLatenciesTimeWasmTaTimeMean{13}

 \def\bmLatenciesRoundtripEnterMean{86}

\def\bmLatenciesRoundtripLeaveMean{20}

 \def\bmLaunchTimeWasmInitRatio{16}
\def\bmLaunchTimeWasmLoadRatio{73}
\def\bmLaunchTimeMemoryRatio{5}
\def\bmLaunchTimeHashRatio{4}
 \def\bmPolybenchTeeGlobalSlowdown{1.34}

 \def\bmInterVsAotGlobalSpeedUp{28}
 \def\bmSpeedtestOneGlobalSlowdownTeeWasm{2.12}
\def\bmSpeedtestOneGlobalSlowdownTeeNative{1.31}
\def\bmSpeedtestOneGlobalSlowdownReeWasm{2.1}
\def\bmSpeedtestOneGlobalRatioTeeWasmAndTeeNative{1.61}
\def\bmSpeedtestOneInsertAverage{2.23}

\def\bmSpeedtestOneSelectAverage{2.04}

 \def\bmAttesterMemMessageZeroMeanInMicrosec{7}

\def\bmAttesterMemMessageOneMeanInMicrosec{50}

\def\bmAttesterMemMessageTwoMeanInMicrosec{5}

\def\bmAttesterKeygenMessageZeroMeanInMs{236}

\def\bmAttesterKeygenMessageOneMeanInMs{235}

\def\bmAttesterSymCryptoMessageOneMeanInMicrosec{88}

\def\bmAttesterSymCryptoMessageTwoMeanInMicrosec{79}

\def\bmAttesterAsymCryptoMessageOneMeanInMs{159}

\def\bmAttesterAsymCryptoMessageTwoMeanInMs{238}
\def\bmVerifierMemMessageZeroMeanInMicrosec{52}

\def\bmVerifierMemMessageOneMeanInMicrosec{7}

\def\bmVerifierMemMessageTwoMeanInMicrosec{7}

\def\bmVerifierKeygenMessageZeroMeanInMs{471}

\def\bmVerifierSymCryptoMessageOneMeanInMicrosec{85}

\def\bmVerifierSymCryptoMessageTwoMeanInMicrosec{80}

\def\bmVerifierAsymCryptoMessageOneMeanInMs{236}

\def\bmVerifierAsymCryptoMessageTwoMeanInMs{159}
\def\bmMessageOneAsymVsSymRatio{2774}
 \def\bmMessageThreeMinimumTime{3}
\def\bmMessageThreeMaximumTime{17}
\def\bmMessageThreeTimeAtOneMB{5.8}
 \def\bmGenannCollectQuoteMean{239}
\def\bmGenannNetHandshakeMean{1.34}
\def\bmGenannSendQuoteMean{1}
\def\bmGenannSumOfWasiRaMin{1.75}
\def\bmGenannSumOfWasiRaMax{1.79}
\def\bmGenannSumOfWasiRaWithoutData{1.58}

\def\bmGenannReceiveDataMinMeanInMs{168}
\def\bmGenannReceiveDataMaxMeanInMs{209}

\def\bmGenannTrainingSpeedUpPercent{1.4}
 
\def\isExtendedCiteEnabled{1}

\ifnum\isExtendedCiteEnabled=1
  \newcommand{\extcite}[2][null]{~\cite{#2}}\else
  \newcommand{\extcite}[2][null]{\def\paramOne{#1}\def\paramNull{null}\ifx\paramOne\paramNull \else ~\cite{#1}\fi }
\fi

\def\isExtendedPaperEnabled{0}

\newcolumntype{R}{>{\raggedleft\arraybackslash}X}
\newcolumntype{P}[1]{>{\raggedleft\arraybackslash}p{#1}}

\newcommand{\sys}{\textsc{WaTZ}\xspace}
\newcommand{\defeq}{\vcentcolon=}
\newcommand{\microsec}{\,{\textmu}s\xspace}
\newcommand{\ms}{\,ms\xspace}
\newcommand{\s}{\,s\xspace}
\newcommand{\Ding}[1]{\raisebox{-0.8pt}{\ding{\the\numexpr #1 + 191}}}
\newcommand{\DingBlack}[1]{\raisebox{-0.8pt}{\ding{\the\numexpr #1 + 201}}}
\newcommand{\polybench}{PolyBench/C\xspace}
\newcommand{\wasm}{Wasm\xspace}
\newcommand{\armarch}{ARM64\xspace}

\colorlet{shadecolor}{Azure2}

\def\BibTeX{{\rm B\kern-.05em{\sc i\kern-.025em b}\kern-.08em
    T\kern-.1667em\lower.7ex\hbox{E}\kern-.125emX}}
\makeatletter
\begin{document}

\title{\sys: A Trusted WebAssembly Runtime Environment with Remote Attestation for TrustZone}

\author{\IEEEauthorblockN{Jämes Ménétrey\,\orcidlink{0000-0003-2470-2827}, Marcelo Pasin\,\orcidlink{0000-0002-3064-5315}, Pascal Felber\,\orcidlink{0000-0003-1574-6721}, Valerio Schiavoni\,\orcidlink{0000-0003-1493-6603}}
\IEEEauthorblockA{University of Neuchâtel, Switzerland --- \emph{first.last@unine.ch}}
}

\maketitle

\def\confname{42nd IEEE International Conference on Distributed Computing Systems (ICDCS'22)}
\def\confyear{2022}
\def\confdoi{10.1109/ICDCS54860.2022.00116}

\definecolor{yellowPaper}{HTML}{fff8ae}
\AddToShipoutPictureFG*{\AtTextUpperLeft{\begin{tcolorbox}[width=\textwidth,colback=yellowPaper,enhanced,frame hidden,sharp corners]  
        \centering\scriptsize
        \copyright~\confyear\ IEEE. Personal use of this material is permitted. Permission from IEEE must be obtained for all other uses, in any current or future media, including reprinting/republishing this material for advertising or promotional purposes, creating new collective works, for resale or redistribution to servers or lists, or reuse of any copyrighted component of this work in other works.
        This is the author's version of the work. The definitive version has been published  in the proceedings of the\\
        \confname.
        \href{https://doi.org/\confdoi}{DOI: \confdoi}
     \end{tcolorbox}   
  }}

\hypersetup{
    pdfcopyright={\copyright~\confyear\ IEEE. Personal use of this material is permitted. Permission from IEEE must be obtained for all other uses, in any current or future media, including reprinting/republishing this material for advertising or promotional purposes, creating new collective works, for resale or redistribution to servers or lists, or reuse of any copyrighted component of this work in other works.
    This is the author's version of the work. The definitive version has been published in the proceedings of the \confname.}
}

\begin{abstract}
    WebAssembly (Wasm) is a novel low-level bytecode format that swiftly gained popularity for its efficiency, versatility and security, with near-native performance.
    Besides, trusted execution environments (TEEs) shield critical software assets against compromised infrastructures.
    However, TEEs do not guarantee the code to be trustworthy or that it was not tampered with.
    Instead, one relies on remote attestation to assess the code before execution.
    This paper describes \sys, which is
    \emph{(i)} an efficient and secure runtime for trusted execution of Wasm code for Arm's TrustZone TEE, and
    \emph{(ii)} a lightweight remote attestation system optimised for Wasm applications running in TrustZone, as it lacks built-in mechanisms for attestation.
    The remote attestation protocol is formally verified using a state-of-the-art analyser and model checker.
    Our extensive evaluation of Arm-based hardware uses synthetic and real-world benchmarks, illustrating typical tasks IoT devices achieve.
    \sys's execution speed is on par with Wasm runtimes in the normal world and reaches roughly half the speed of native execution, which is compensated by the additional security guarantees and the interoperability offered by Wasm.
    \sys is open-source and available on GitHub along with instructions to reproduce our experiments.{\parfillskip=0pt\par}
\end{abstract}

\section{Introduction}

Security is critical when designing and deploying distributed applications for mutually distrusting stakeholders, \eg hosting companies. The problem grows in complexity in heterogeneous systems, when one considers decentralised operations of IoT, edge and cloud devices, all at risk of being compromised.

Trusted execution environments (TEEs), \eg Intel SGX~\cite{cryptoeprint:2016:086} and Arm TrustZone~\cite{10.1145/3291047}, offer hardware support for securely executing applications in shielded environments.
While TEEs are promoted by the commercial offerings of major cloud providers, they do not guarantee that the code itself is trustworthy and has not been tampered with.
Remote attestation\extcite[7544334]{nunes2019vrased,7544334,eldefrawy2012smart,ibrahim2017seed,koeberl2014trustlite} is typically used to assess the code before its execution.
However, while this key security feature is provided by some TEEs (\eg Intel SGX\extcite{sardar2020towards}), Arm TrustZone lacks built-in support for remote attestation.
Given the swift growth of popularity of Arm-based architectures in the IoT edge computing\extcite{stateofiot} and its recent adoption in the general computer market\extcite{applem1}, this is concerning.
In addition, recent attacks have shown how to compromise IoT devices via malicious firmware updates~\cite{9249434} or software flaws~\cite{7932855}, typically beyond the threat model of TrustZone.
Nonetheless, the inability of manufacturers to verify the authenticity of their software upon execution makes these platforms inadequate for handling sensitive tasks and data, in particular when dealing with recent scenarios such as trustworthy machine learning systems at the edge~\cite{li2018learning}.

WebAssembly (\wasm)~\cite{10.1145/3062341.3062363} is a new bytecode standard for running applications at near-native speed.
It enables developers to build their software components with the productivity benefits of modern programming languages while supporting legacy code, since modern compilers support \wasm~\cite{10.5555/977395.977673}.
TrustZone is a constrained environment that runs small executables using a specialised API. \wasm fits well in this model, thanks to its small runtime overhead, portability, supporting non-standard system interfaces, and fast as the bytecode can be compiled just-in-time and ahead-of-time.

Embedding \wasm in TrustZone can be helpful in many use cases.
The smartphone industry could rely on \wasm as an interoperable bytecode to execute secure applications inside TrustZone, as exclusively reserved for the manufacturer's needs or for strategic partners~\cite{teegris}.
Automotive applications could rely on IoT devices to run machine learning algorithms inside enclaves, ensuring the validity of the results via attestation and using \wasm to leverage legacy machine learning frameworks.
We demonstrate the latter in \S\ref{sec:macro-ra}.

We present \sys, an efficient and secure runtime for trusted execution of \wasm code inside TrustZone, adding support for remote attestation.
We leverage the sandbox isolation of \wasm to mitigate, and possibly prevent, vertical privilege escalations and lateral attacks. 
We combine TrustZone with \wasm bytecode to issue trustworthy evidence, attesting the genuineness of running software.
In \S\ref{sec:ra}, we adapted and fully implemented the remote attestation protocol from SGX~\cite{intelraendtoend}, using a public-key infrastructure.
As such, we facilitate the deployment of fully decentralised applications spanning various devices at the core or the edge of the network.
\sys extends OP-TEE~\cite{optee}, a popular open-source trusted OS, and we validated our prototype with Arm hardware.

The main research questions that \sys intends to answer, and the main contributions of this work are the following.

\begin{shaded*}
    \textit{\textbf{RQ1}: Are there system challenges when embedding \wasm into Arm TrustZone?}
\end{shaded*}

TrustZone requires using a trusted operating system, which declares a non-standardised API to interact with system resources.
This, coupled with the constrained nature of TEEs (\eg no system call, limited memory), increases the complexity of hosting general-purpose applications compiled in \wasm.
We show (\S\ref{sec:overview}) that \sys is the first system to run \wasm applications in TrustZone, while leveraging the WASI standard, a POSIX-like layer for \wasm, to interact with the TEE facilities. 
As such, hosted applications seamlessly communicate with TrustZone, while abstracting from that trusted environment's specific API.
This software encapsulation strategy for TrustZone enhances the state of the art: each software deployed in the trusted world is fully isolated and cannot interfere with others, similarly to SGX enclaves, thanks to the robust sandbox provided by \wasm. 
Our \sys prototype supports a subset of WASI API to run general-purpose software and to evaluate a database engine (SQLite~\cite{5231398}), as well as a machine learning library (Genann~\cite{genann}).
We plan to widen the WASI support for additional system calls, \eg, file system and networking.

\begin{shaded*}
    \textit{\textbf{RQ2}: How can relying parties trust the remote execution of \wasm applications?}
\end{shaded*}

Given the lack of built-in remote attestation in TrustZone, we propose a protocol (\S\ref{sec:ra}) to attest \wasm code embedded in our trusted environment.
Hence, we ensure executed \wasm binaries are trustworthy, similarly to Intel SGX.
We identify the hardware requirements of IoT devices to provide attestable guarantees (\eg root of trust, secure boot), showing how to be combined with the trusted environment to verify \wasm binaries.
We contribute an extension of the WASI specifications, called WASI-RA, to enable the hosted \wasm applications to attest against trusted parties and communicate shared secrets and confidential data, based on Intel SGX's remote attestation protocol.
Finally, we demonstrate that \sys is an end-to-end solution that leverages the fingerprint (\ie a hash of the bytecode) of isolated \wasm applications, hardware elements and the attestation mechanism to induce trust into IoT devices.

\begin{shaded*}
    \textit{\textbf{RQ3}: How does the performance of \wasm applications compare when hosted in the trusted environment?}
\end{shaded*}

We extensively evaluate \sys (\S\ref{sec:evaluation}) against micro- and macro-benchmarks, together with SQLite and Genann.
Performance is on par with \wasm applications executed outside of the TEE, and up to \bmSpeedtestOneGlobalSlowdownTeeWasm$\times$ compared to native execution, deemed negligible compared to the benefits \sys offers.
We contribute and describe an extension of the trusted kernel (OP-TEE) to execute ahead-of-time compiled \wasm applications to achieve these performance results.
Besides, we perform a security analysis of \sys (\S\ref{sec:security}), as well as formal verification of the remote attestation protocol.

\smallskip\noindent\textbf{Roadmap.}\quad
The remainder of the paper is organised as follows.
We first present background information and related work in \S\ref{sec:background}.
\S\ref{sec:overview} introduces the overall design and architecture of our \sys runtime.
The remote attestation mechanism of \sys is described in \S\ref{sec:ra}.
We elaborate on some implementation details in \S\ref{sec:implementation} and present our extensive evaluation of \sys in \S\ref{sec:evaluation}.
We analyse the security of our approach in \S\ref{sec:security}, before concluding in \S\ref{sec:conclusion}.

\section{Background and related work}
\label{sec:background}

Our \sys runtime supports trusted execution of \wasm code inside TrustZone with remote attestation mechanisms.
This section briefly introduces the underlying technologies and highlights how our approach improves related work.

\smallskip\noindent\textbf{Arm TrustZone.}\quad It provides the hardware elements enabling TEEs on Arm processors~\cite{tz-explained}.
TrustZone enables a single TEE per system, called secure world, as opposed to the normal world, the untrusted environment.
The CPU lives between two security states: secure and normal worlds, switching via a secure monitor instruction (\texttt{SMC}).
System resources are strictly isolated: the normal world cannot access the resources (\eg memory, peripherals, \etc) reserved for the secure world.
During the \emph{secure bootstrap} of the secure world, an integrity check of its software image establishes a chain of trust.

OP-TEE~\cite{optee} is a popular open-source runtime environment with native support for TrustZone.
It offers a developer-friendly setup~\cite{gottel2019developing} to build trusted applications.
OP-TEE follows the TEE architecture and API standardised by GlobalPlatform (GP API)~\cite{gp-online}, built around three components: a client application, a dedicated Linux driver and the OP-TEE OS.
The OS of the normal world is referred to as \emph{rich execution environment} (REE).
The host application runs in the normal world, as a client of a trusted application (TA) in the secure world.
Host applications leverage \emph{client} APIs.

\smallskip\noindent\textbf{WebAssembly.}\quad \wasm~\cite{10.1145/3062341.3062363} is a W3C open standard for a portable, compact, low-level, stack-based binary code format.
Initially intended for building browser applications, the specifications allow for standalone execution running outside browsers.
We exploit the \wasm's sandbox capabilities, including software fault isolation~\cite{10.1145/168619.168635} and control-flow integrity~\cite{10.1145/1102120.1102165} to isolate \wasm code from the trusted OS.
We also leverage the WebAssembly system interface (WASI)~\cite{wasi}, a POSIX-like interface used by \wasm programs to interact with the underlying OS.
WASI acts as a mediator between the \wasm application and the GP API.
As a result, many programming languages (\eg C/C++, Rust, Swift, Go) can target \wasm with WASI so they require no modification to run as \wasm binaries.

\smallskip\noindent\textbf{WebAssembly and TEEs.}\quad
Few options exist to host \wasm applications inside TEEs.
\textsc{Twine}~\cite{twine}, an embedded trusted runtime for WebAssembly, executes \wasm applications inside Intel SGX enclaves.
\textsc{Twine} relies on the WebAssembly micro runtime (WAMR)~\cite{wamr}, enabled with WASI to interact with a secure file system.
Enarx~\cite{enarx} targets Intel SGX enclaves and AMD SEV virtual machines.
Veracruz~\cite{veracruz} only supports VM-based TEEs, such as Arm CCA~\cite{armcca} and AWS Nitro~\cite{awsnitro} enclaves, having recently dropped SGX and TrustZone enclaves considered too constraining~\cite{veracruzDroppedSupport}.
AccTEE~\cite{10.1145/3361525.3361541} and Se-Lambda~\cite{10.1007/978-3-030-01701-9_25} run \wasm binaries in Intel SGX enclaves using the V8 JavaScript/\wasm engine.
AccTEE provides trusted resource accounting, while Se-Lambda deploys serverless programs over function-as-a-service.
Contrary to the mentioned solutions, \sys runs on Arm TrustZone, leverages WASI for system interactions and supports trustworthy code execution, thanks to the close integration with \wasm and the remote attestation mechanism.
We propose a functional paradigm for Arm TrustZone: every hosted Wasm application is isolated from the rest of the trusted world via the \wasm sandbox.
We note that OP-TEE requires every TA to be signed to be trusted and executable in the trusted world.
This is a significant impediment when offering trusted execution for third parties, which is solved with \sys.
This approach maintains the security of TrustZone thanks to the isolation of \wasm sandbox.

\newcommand{\YES}{\textcolor{NavyBlue}{\ding{51}}}
\newcommand{\NO}{\color{BrickRed}{\ding{55}}}
\newcommand{\requireadaptation}{\textsuperscript{*}}
\newcommand{\theresnorust}{\textsuperscript{\textdagger}}
\begin{table}[!t]
\centering
\small
\setlength{\tabcolsep}{4pt}
\rowcolors{1}{gray!10}{gray!0}
  \begin{tabularx}{\columnwidth}{X@{}ccccccc}
  \toprule
  \rowcolor{gray!25}
                    &AOT    & WASI          & RA      & \makecell[c]{RA in\\WASI}  & $\mu$RT                            & \makecell[c]{IoT\\TEE}    & TEE(s)\\
  \midrule   
  \textsc{Twine}    & \YES                      & \YES          & \NO     & \NO                         & \YES                            & \NO                       & SGX\\
  Veracruz          & \NO                       & \YES          & \YES    & \NO                         & \NO                             & \NO                       & Nitro, CCA\\
  Enarx             & \NO                       & \YES          & \YES    & \NO                         & \NO                             & \NO                       & SGX, SEV\\ AccTEE            & \NO                       & \NO           & \NO     & \NO                         & \NO                             & \NO                       & SGX\\ Se-Lambda         & \NO                       & \NO           & \YES    & \NO                         & \NO                             & \NO                       & SGX\\ Teaclave          & \NO                       & \NO           & \YES    & \NO                         & \YES                            & \NO                       & SGX\\ \midrule
  \sys              & \YES                      & \YES          & \YES    & \YES                        & \YES                            & \YES                      & TrustZone\\
  \bottomrule
  \end{tabularx}
  \caption{\label{tab:rw-comparison}Comparison of the related work features.}
  \vspace{-12pt}
\end{table}

\autoref{tab:rw-comparison} compares \sys against state-of-the-art TEE runtimes for Wasm along the following dimensions: \emph{AOT} (process ahead-of-time compiled \wasm bytecode), \emph{WASI} (enable system interaction), \emph{RA} (support remote attestation), \emph{RA in WASI} (provide a WASI API to control the remote attestation in the hosted \wasm application), \emph{$\mu$RT} (use a small runtime, less than 1\,MB in memory), \emph{IoT TEE} (designed for IoT devices), and \emph{TEE(s)} (summarises the TEE technologies).

\smallskip\noindent\textbf{Remote attestation (RA).} It verifies the genuineness of software in a remote process.
It relies on a challenge-response protocol~\cite{ietf-rats-architecture-12}.
An \emph{attester} must prove the software's uncompromised state. It shares \emph{evidence} (a \emph{quote} in the SGX jargon) to a \emph{verifier}, which assesses its genuineness.
The evidence is a cryptographically signed proof of \emph{claims}, \ie pieces of asserted information, such as a code measurement hash.
The verifier is configured with \emph{references values}, compared against the received claims for validation.
Finally, \emph{endorsements} are secure statements used by the verifier to identify the devices eligible to issue trustworthy evidence.

Remote attestation can be based on software, hardware or a combination of both~\cite{7544363, 8342055}.
Recent work addresses the remote attestation mechanisms and protocols, including for IoT and edge devices~\cite{sota-dais,7544334,197162,10.1145/3319535.3363205,Ahn2020DesignAI,chen2020mage,10.1145/3098954.3098971}.
Intel SGX has built-in support for remote attestation~\cite{anati2013innovative}.
\sys follows the same principles to deliver RA on code executed on Arm processors.
Ling \emph{et al.}~\cite{LING2021102240} proposed a trusted boot mechanism with remote attestation of software running in Arm's normal world.
Similarly to \sys, they leveraged OP-TEE to ensure the boot and runtime state of IoT devices are trustworthy.
However, their approach depends on the integrity of the normal world kernel.
Our proposal does not rely on a dependable OS in the normal world: the attested software runs isolated in the secure world.
\sys extends the design of Intel's remote attestation protocol to use software enclaves created in the trusted world, through isolation within \wasm sandboxes.
\sys is strongly integrated with \wasm by providing WASI-RA.
As such, it is the first system enabling the hosted \wasm applications to control the remote attestation process.
\sys measures the \wasm bytecode that bootstraps the enclave to issue evidence.
As a result, third parties will remotely appraise software and authenticate IoT edge devices by leveraging TrustZone, a hardware-based root of trust and secure boot, to build a hardware-enforced attesting environment.

\vspace{-2pt}
\section{\sys: System overview}
\label{sec:overview}
\vspace{-2pt}

This section introduces \sys's threat model, design, architecture, and details of the trusted runtime.

\smallskip\noindent\textbf{Threat model}.\quad
We consider the following aspects.

\emph{(a) Hardware.}
\sys leverages the following hardware capabilities:
\begin{enumerate*}[label=\emph{(\roman*)}]
    \item TrustZone security extensions,
    \item a root of trust, and
    \item secure boot.
\end{enumerate*}
While we consider a powerful attacker with physical access to the devices, we assume that the protections offered by hardware cannot be subverted.
As such, the adversary may fully control the system on a chip (SoC) as well as its peripherals, but excluding all the components belonging to TrustZone.
Consequently, \sys cannot defend itself from physical attacks to the volatile memory space assigned to TrustZone, which is not encrypted, unlike Intel SGX.
We also consider storage rollback attacks out of scope, which can be mitigated using hardware monotonic counters~\cite{martin2021adam}.

\emph{(b) Secure world.}
We assume the secure monitor, bootloader and trusted OS do not contain vulnerabilities enabling an attacker to breach the TEE.
The cryptographic primitives and algorithms are considered correct.
Code and data inside the TEE are trusted and cannot be accessed from the normal world, except through dedicated channels controlled by \sys.
Finally, side-channel attacks~\cite{10.1145/3319535.3354197,10.1145/3319535.3354201,229832,8486293,spectre} are out of scope.

\emph{(c) Normal world.}
We make no assumption regarding the normal world, which includes the rich OS and the user space.
Compromised OSes may arbitrarily respond to trusted OS calls, causing its malfunction.
The trusted applications relying on the normal world should be carefully crafted to ignore abnormal responses or even abort execution in such cases.

\begin{figure}[!t]
    \centering
    \includegraphics[scale=0.5]{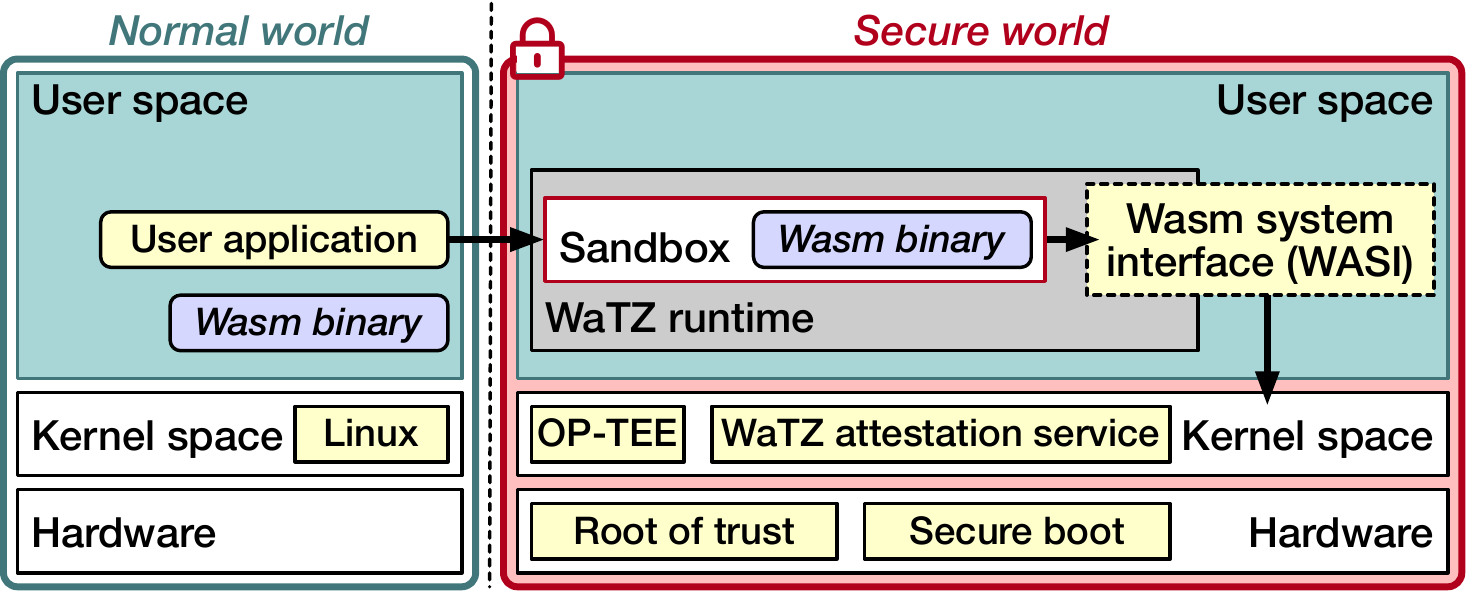}
    \caption{Overall architecture of \sys.}
    \label{fig:runtime:arch}
    \vspace{-10pt}
\end{figure}

\smallskip\noindent\textbf{Design overview.}\quad \sys is a trusted runtime to execute \wasm applications with remote attestation capabilities.
\autoref{fig:runtime:arch} illustrates its components.
Its small footprint (265\,kB on disk, including the runtime and \sys's components) brings several advantages.
Smaller programs offer smaller attack surfaces.
TEEs and small devices (edge or IoT) are usually tight in memory.
Its small footprint allows deploying and running a complete \wasm virtual machine inside TrustZone using a small edge-scale Arm processor.
We estimate that the increase of the trusted computing base (TCB) due to the embedding of the \wasm runtime in the TEE is outweighed by its benefits.

To provide trusted OS features to \wasm applications, we contribute an adaptation layer that binds WASI to the API existing in the trusted environment (GP API in our prototype).
This, in conjunction with the reliance of \wasm applications on WASI, allows \sys to run unmodified \wasm applications, while benefiting from a TEE.
While we support a subset of the WASI API enabling to execute general-purpose software, the adaptation layer is effortlessly extensible to support additional system calls, such as file system and networking.

To offer remote attestation, we designed and implemented WASI-RA, an extension of WASI enabling the hosted Wasm applications to interact with the process of attestation.
We guarantee that \sys is booted correctly and was not tampered with using a secure boot system.
When \sys loads a \wasm application, its bytecode is stored in the secure memory and measured to produce a hash.
Using our interface, a \wasm attester can request \sys to generate trusted evidence based on the hardware secret and the hash of the \wasm bytecode.
Further, \sys includes a remote attestation protocol to enable a third-party verifier to check if evidence is genuine, which relies on the measured fingerprint of the \wasm applications.
Upon positive attestation, our protocol simplifies the establishment of a hybrid cryptosystem for the verifier and the attester, which can be later used to create secure channels.

\smallskip\noindent\textbf{Embedded runtime.}\quad
We built \sys as an embedded \wasm runtime with a WASI interface as this design provides several advantages, \ie removing some barriers for building TAs.
First, \sys opens the choice for programming languages.
Provided that the compiler can emit \wasm bytecode and supports WASI, it is a clear advantage over vanilla OP-TEE, which limits developers to C only.
Second, this hides the complexity of writing code dedicated to OP-TEE since WASI calls are generated by the compiler and they abstract the implementation details of GP API.
Furthermore, WebAssembly separates the virtual address spaces used for \wasm applications and the native runtime process, and code, stacks and heap are handled separately, making memory-oriented attacks or developer mistakes more unlikely.
Besides, WASI ensures that the applications do not harm the secure world and acts as a gatekeeper to run operations outside of the runtime.
Finally, applications are not tightly coupled to the underlying TEE, and \sys can load any regular \wasm/WASI application without changes.
As a result, \wasm brings more flexibility, versatility and security compared to their native counterpart, allied with its strong sandboxing mechanism that isolates each hosted \wasm application.
This newly introduced isolation layer extends the single trusted world of TrustZone to an environment with multiple secure and mutually distrusting enclaves, similarly to the isolation scheme offered by Intel SGX.
Hence, \sys changes the paradigm of software deployment and execution of the secure world to a more relaxed approach, which previously required a trusted application to be signed, while not introducing any security drawback.

Instances of trusted \wasm applications are started by a user space process in the normal world, which uses the standard OP-TEE API to prepare a buffer containing the \wasm application and trigger \sys in the secure world.
Once in the secure world, \sys copies the bytecode into a secure, sandboxed memory, calculates the hash for future attestation and starts the execution immediately.
\wasm applications natively use the standard WASI interface to interact with the OS, which, in turn, diverts the calls to \sys.
We implemented the \sys attestation service as a new kernel module for OP-TEE, enabling the runtime to generate evidence for relying parties to prove the applications' authenticity.
While several open-source alternatives exist to execute \wasm code, we settled for WAMR~\cite{wamr}, a lightweight and embeddable runtime implemented in C, ideal for TEEs in general (small TCB) and OP-TEE in particular, since TAs are written in C.

\smallskip\noindent\textbf{Execution modes.}\quad
WAMR can execute code in three modes, each with its benefits and drawbacks: interpreted, compiled just-in-time (JIT) and compiled ahead-of-time (AOT).
Interpreted is the simplest yet slowest, as it does not require pre-processing the bytecode.
When using JIT compilation, the bytecode is translated into native code whenever executed, but embedding a compiler in the runtime increases its complexity, size and dependencies.
Indeed, WAMR uses LLVM~\cite{10.5555/977395.977673} as its JIT and AOT compiler, which is not trivial to port to a restricted environment like OP-TEE.
With AOT compilation, the bytecode is translated before execution, so the runtime does not need to include a compiler, but requires the TA to allocate executable memory.
We opted for AOT compilation for \sys's runtime.
However, OP-TEE's memory management API cannot modify the pages' protection to mark them as executable~\cite{opteeexecutablecode}.
Hence, we extended the trusted kernel to provide such capabilities to TAs.
The AOT execution speed is on average  \bmInterVsAotGlobalSpeedUp$\times$ faster than with interpretation.

\section{Remote attestation of WebAssembly}
\label{sec:ra}

This section presents the remote attestation mechanism, first by 
explaining how the hardware is trusted, extending this principle to the secure OS and \sys.

\smallskip\noindent\textbf{Root of trust.}\quad
We designed \sys for devices that expose a hardware root of trust to the secure world.
We extended OP-TEE to deterministically derive a key pair from the hardware root of trust.
Normal and secure OS can hence be updated without losing the key materials, and on-chip key generation guarantees that the private key never leaves the trusted kernel OS.
The public key is then exported and used as an endorsement value to be verified during remote attestation requests.
This key pair, called \emph{attestation keys}, is at the core of \sys's mechanisms to provide attestable signatures and guarantee platform authenticity.

\smallskip\noindent\textbf{Secure boot.}\quad
Secure boot is a security mechanism to ensure that the device is booting in a trusted state.
\sys requires the device to implement secure boot, so only trusted entities are able to provision software to boot the secure world (a couple of bootloaders and the trusted OS).
Therefore, this enforces a \emph{chain of trust} that protects the attestation keys against extraction from the secure kernel OS.
The boot sequence is as follows: the first-stage bootloader (ROM) verifies if the second-stage bootloader is genuine, based on the public key stored in one-time programmable fuses (eFuses)~\cite{imx8securityreference}.
The previous booting component recursively verifies the next boot stages until the secure world is fully booted.

\smallskip\noindent\textbf{Proof of trust: the evidence.}\quad
\label{sec:evidence}
\sys generates cryptographically signed reports, called \emph{evidence}, asserting that an executing \wasm application is trustworthy and the device genuine, by producing a hash of the \wasm AOT bytecode stored in the secure memory at launch time.
We offer an API to \wasm applications (see WASI-RA in \S\ref{sec:implementation}) to issue evidence and establish a secure communication channel with a verifier.
Then, the evidence is checked by the verifier using the corresponding public key of the device and examines the code measurement to match its reference values.

The evidence is created by interacting with the attestation service, implemented as a kernel module in OP-TEE (shown in \autoref{fig:runtime:arch}).
The evidence includes 
\begin{enumerate*}[label=\emph{(\roman*)},noitemsep,nolistsep,leftmargin=*]
    \item an anchor, which is a value defined by the transport layer to bind security parameters to a particular session (\eg a public session key),
    \item the version of \sys, enabling the relying party to exclude outdated systems,
    \item the claim, \ie the bytecode hash,
    \item the public key of the attestation service, for the verifier to determine if the device is endorsed, and
    \item the digital signature of the evidence.
\end{enumerate*}

\smallskip\noindent\textbf{Security requirements.}\quad
Our remote attestation protocol satisfies a number of security requirements as specified below:
\begin{enumerate}[noitemsep,nolistsep,leftmargin=*] \item \emph{Mutual key establishment}: A shared secret key is established for communication between the attester and the verifier, using the elliptic-curve Diffie–Hellman ephemeral (ECDHE) key-agreement protocol.
    \item \emph{Mutual entity authentication}: The attester and the verifier are mutually authenticated to prevent masquerading attacks.
    From the attester's standpoint, the verifier's public key must be hardcoded into the \wasm application.
    This, combined with the application measurement, ensures that an attacker cannot change the key so that the software can only communicate with the intended remote service.
    \item \emph{Half trust assurance}: The attester attests the \wasm application and the platform integrity to the verifier.
    The verifier does not provide a similar proof to the attester, and the attester assumes the entity authentication is sufficient.
    \item \emph{Freshness}: ECDHE (\emph{ephemeral}) requires the key pairs to be fresh, hence preventing replay attacks.
    \item \emph{Forward secrecy}: Compromised long-term secrets do not affect the security of earlier or future exchanges.
    Similarly to \emph{freshness}, ECDHE achieves this goal, which means the keys are renewed for every tentative of remote attestation.
\end{enumerate}

\smallskip\noindent\textbf{\sys protocol for remote attestation.}\quad
\label{lab:trusted-runtime:ra}
The GP API defines an interface to establish a secure communication channel using TLS.
However, OP-TEE lacks the corresponding implementation~\extcite[opteefaq]{globalplatformtls,opteefaq}.
We extended and implemented the RA protocol of Intel SGX~\cite{intelraendtoend} (itself inspired by SIGMA~\cite{10.1007/978-3-540-45146-4_24}) to not rely on TLS.
We changed the protocol compared to the original in various aspects:
\begin{enumerate*}[label=\emph{(\roman*)}]
    \item removed the SGX specificities, such as the interaction with the quoting enclave, as the kernel module of \sys provides the measurements,
    \item merged the two first messages to communicate from the client to the server as they tightly relate,
    \item provided a fixed structure for the last message to seamlessly handle confidential data, eliminating the burden of a hosted \wasm application from decrypting that content,
    \item omitted Intel's SGX Enhanced Privacy ID (EPID) for conciseness, and
    \item removed the dependency on Intel's PKI, since the device's key pair is emitted by \sys based on the embedded root of trust.
\end{enumerate*}
\autoref{tab:ra-protocol} formalises the remote attestation protocol.
Below, we detail each protocol's message and the required cryptographic operations.

\smallskip\noindent{\emph{(a) Message 0 (attester\textrightarrow verifier)}}:
The attester generates a session key pair $<a, G_a>$ and sends the public part $G_a$.

\smallskip\noindent{\emph{(b) Message 1 (attester\textleftarrow verifier)}}:
Upon reception of $\text{msg}_0$, the verifier generates a session key pair $<v, G_v>$.
It computes the shared secret from the public session key of the attester $G_a$ and its private session key $v$, which gives $G_{av}$.
This shared secret is derived into a \emph{key derivation key} (KDK), which is further derived into two shared secrets: $K_m$ for calculating MACs and $K_e$ for future messages encryption in the session.
These derivations are the same as in Intel SGX~\cite{intelraendtoend}.
A reply message is sent to the attester, containing $G_v$, the verifier's ECDSA public key $V$ (its identity), and a signature of both public session keys.
The message is appended with a MAC.

\newcommand{\concat}{\:\|\:}
\setlength{\belowdisplayskip}{2pt}
\setlength{\abovedisplayskip}{2pt}
\setlength{\belowdisplayshortskip}{2pt}
\setlength{\abovedisplayshortskip}{2pt}
\begin{table}[!t]
\centering
\small
\setlength{\tabcolsep}{4pt}
    \begin{tabularx}{\columnwidth}{lclcX}
    \toprule
    \rowcolor{gray!25}
    $\text{msg}_0$      & $\defeq$              & \multicolumn{3}{l}{$G_a$} \\
    $\text{msg}_1$      & $\defeq$              & \multicolumn{3}{l}{$\text{content}_1 \concat \text{MAC}_{K_m}(\text{content}_1)$} \\
                        & $\hookrightarrow$     & $\text{content}_1$ & $\defeq$ & $G_v\concat V \concat \text{SIGN}_V(G_v \concat G_a)$ \\
    \rowcolor{gray!25}
    $\text{msg}_2$      & $\defeq$              & \multicolumn{3}{l}{$\text{content}_2 \concat \text{MAC}_{K_m}(\text{content}_2)$} \\
    \rowcolor{gray!25}
                        & $\hookrightarrow$     & $\text{content}_2$ & $\defeq$ & $G_a \concat \text{evidence} \concat \text{SIGN}_A(\text{evidence})$ \\
    \rowcolor{gray!25}
                        & $\hookrightarrow$     & $\text{evidence}$ & $\defeq$ & $(\text{anchor} \concat A \concat \hdots)$\\
    \rowcolor{gray!25}
                        & $\hookrightarrow$     & $\text{anchor}$ & $\defeq$ & $\text{HASH}(G_a \concat G_v)$\\
    $\text{msg}_3$      & $\defeq$              & \multicolumn{3}{l}{$iv \concat \text{AES-GCM}_{K_e}(data)$} \\
    \bottomrule
    \end{tabularx}
    \caption{\label{tab:ra-protocol}The remote attestation protocol of \sys.}
    \vspace{-12pt}
\end{table}

\smallskip\noindent{\emph{(c) Message 2 (attester\textrightarrow verifier)}}:
The attester verifies the signature of the public session keys: different session keys may reveal a masquerading or replay attack, and assesses the MAC of $\text{msg}_1$.
It also checks whether the service public key $V$ matches the hardcoded key in the \wasm application.
Doing so ensures the attester communicates with the intended service and prevents an attacker from altering that key as it is part of the code measurement.
The attester computes the shared secret from the public session key of the verifier $G_v$ and the private session key $a$, which gives $G_{va}$ that is equal to $G_{av}$ computed by the verifier.
The key derivations follow the same process as in $\text{msg}_1$.
The attester creates $\text{msg}_2$ by concatenating its public session key $G_a$ with a newly generated evidence (see \S\ref{sec:evidence} for details) signed by the attester $A$, where the anchor of the transport layer is the hashed concatenation of the public session keys.
Finally, it appends a MAC.

\smallskip\noindent{{\emph{(d) Message 3 (attester\textleftarrow verifier)}}:
The verifier checks the MAC of $\text{msg}_2$ and verifies that $G_a$ matches the one received in $\text{msg}_0$.
It also examines whether the anchor corresponds to the public session keys, revealing a masquerading or replay attack.
It extracts the evidence's public attestation key and checks against its list of endorsed public keys to determine whether this is a known device.
If the key is found, the digital signature of the evidence is checked, which indicates whether the hardware is genuine.
Finally, to verify that the \wasm application is trustworthy, its code measurement claim is compared with a list of possible reference values.
If all verifications pass, the protocol sends $\text{msg}_3$ with an arbitrary confidential data, called \emph{secret blob}, encrypted with AES-GCM, which requires $iv$ (an initialisation vector).

We simplified the protocol by omitting the use of session identifiers.
Such identifiers are needed for having multiple sessions with concurrent remote attestation requests.
We also reduced the complexity by keeping the evidence in clear.
If the secrecy of this structure is a concern, the protocol can be extended to protect the evidence using AES-GCM.
Later, in \S\ref{sec:security:protocol}, we describe how we formally verified this protocol.

\ifnum\isExtendedPaperEnabled=1
\fi

\section{Implementation}
\label{sec:implementation}
This section described the implementation details of \sys.
Note that our prototype requires the following features from the underlying hardware platform:
\begin{enumerate*}[label=\emph{(\roman*)}]
    \item a root of trust,
    \item a secure boot to harden a chain of trust, and
    \item the TrustZone extensions in the CPU.
\end{enumerate*}
As said, \sys relies on OP-TEE, and we detail the extensions in the following.

\smallskip\noindent{\textbf{Overview.}}\quad
\autoref{fig:prototype} illustrates the components of our prototype, which reflects the architecture defined in \S\ref{sec:overview}.
It comprises an attester and a verifier.
Initially (\DingBlack{1}), a \wasm application is loaded by the normal world in the TEE.
\sys runtime copies the bytecode into the secure memory, measures it and executes the application.
Later on, the hosted application can use remote attestation to fetch the secret blob from a relying party. This consists of contacting a verifier to create a secure communication channel (\DingBlack{2}).
An anchor value is associated with the RA handshake, which is forwarded to the attestation service in the trusted kernel to issue evidence (\DingBlack{3}).
Lastly, the verifier checks whether the device is genuine, based on the evidence: if trustworthy, it sends the secret blob using the secure channel (\DingBlack{4}).
In the remainder of this section, we describe further details of the inner working of \sys.

\begin{figure}[!t]
    \centering
    \includegraphics[scale=0.5]{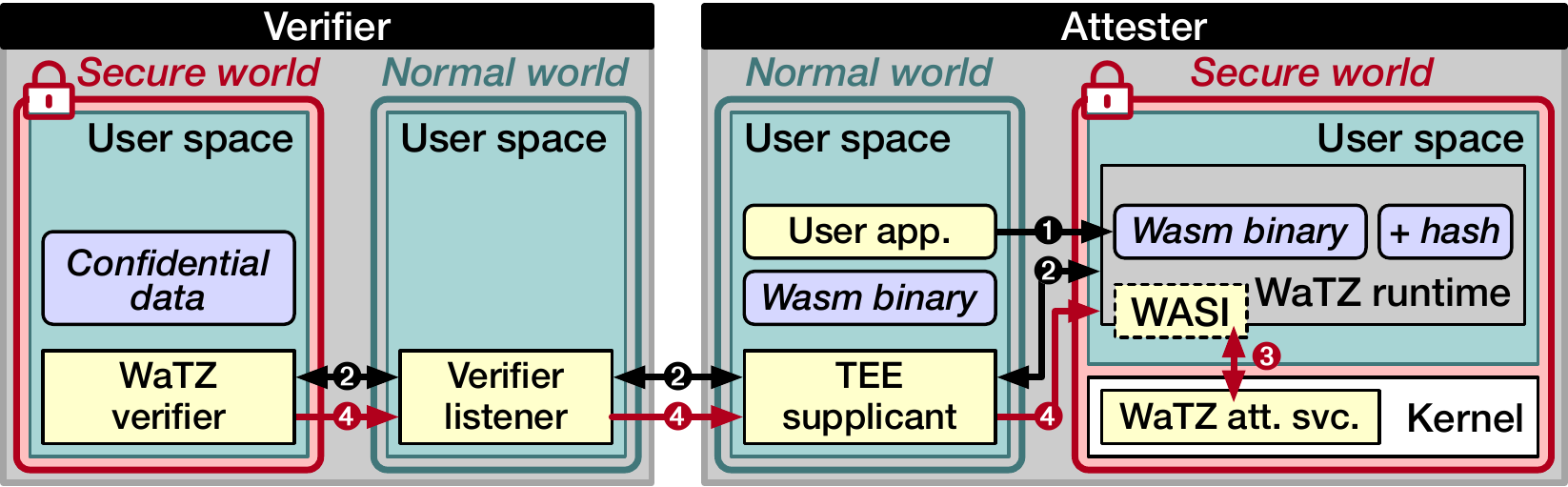}
    \caption{Components of the prototype.}
    \label{fig:prototype}
    \vspace{-10pt}
\end{figure}

\smallskip\noindent{\textbf{The runtime (attester).}}\quad
Our \wasm runtime is a trusted application written in C (1.6\,k SLOC), statically linked to a modified version of WAMR.
As Wasm applications call the OS using WASI, we implemented a WASI interface, mapping its calls to the functions available in the GP API.
To find all the required calls used to execute our experiments properly, we first manually coded dummy functions for all 45 WASI API functions, throwing exceptions when called.
Then, we implemented adapters for the WASI functions necessary for our benchmarks (0.9\,k SLOC) to use whatever was available in GP.
While our experiments could successfully be executed without adding extra features to GP, many standard functionalities (\eg, thread management) are not supported by TAs.
However, we note that with some extra engineering efforts, \sys may be completed to support file system interaction via the Trusted Storage API and support other missing functions leveraging (parts of) an embedded C library or a library kernel. 

The GP API includes TCP/IP sockets, implemented by OP-TEE.
Under the hood, the trusted kernel redirects the communication to the normal world, using a small shared memory buffer to transfer data.
OP-TEE comes with a built-in user space supplicant daemon that runs in the normal world and is responsible for services such as network communication.

When starting a \wasm application, its AOT compiled bytecode is copied by the runtime from the normal world.
The implementation of OP-TEE does not allow TAs to access memory from the normal world directly.
Instead, the normal world allocates a shared memory buffer accessible by both worlds.
OP-TEE limits the amount of memory available for shared buffers. 
We increased the limit to 9\,MB (the largest value that would not break OP-TEE).
A similar problem occurs when allocating memory inside the trusted world, which we modified to allow up to 27\,MB.
Pushing further the memory limits leads to OP-TEE malfunctions.
A quick investigation of its code indicates the likely reasons to be in the data structures used to maintain memory regions, not the Arm architecture itself~\cite{opteefaq}.
According to the GitHub repository, increasing the memory cap of OP-TEE is a recurring request, and workarounds vary depending on the hardware.
We believe that such limits could be removed (or pushed much further away) in future versions of OP-TEE.
As our implementation and experiments did not require substantial amounts of memory, we did not address these constraints.

\wasm applications are loaded by the runtime rather than by the regular native loader of OP-TEE.
As such, \sys is required to allocate executable memory pages to store the AOT compiled bytecode.
We extended OP-TEE to provide such capability by implementing an additional system call similarly to POSIX \texttt{mprotect}.
Consequently, \sys can allocate executable memory pages to run AOT \wasm bytecode, which is \armarch assembly code.
We plan to submit a pull request with this improvement to the OP-TEE upstream repository.

\smallskip\noindent{\textbf{The server (verifier).}}\quad
Our prototype implements a server application that acts as a verifier in the remote attestation protocol.
It comprises two parts: \emph{(i)} the listener in the normal world and \emph{(ii)} the verifier as a TA in the secure world.
While the verifier is currently implemented as a TA written in C (1.4\,k SLOC), any environment that supports the cryptographic operations mentioned in \S\ref{sec:ra} can host this service.
As such, the remote attestation server could be executed in an Intel SGX enclave or as a plain service in a regular OS.

\ifnum\isExtendedPaperEnabled=1
\fi

Contrarily to the attester, the verifier must have a dedicated application in the untrusted side of the TEE because the GP API for sockets lacks the capability of listening for incoming connections.
For that matter, the messages are received in the listener application and forwarded to the TA.
Similarly, the messages generated by the TA are handed to the untrusted application for delivery.
Messages are stored in shared memory buffers, and their handling occurs in the secure memory, acting as the TEE-supplicant provided by OP-TEE.

\smallskip\noindent{\textbf{The attestation service.}}\quad
We designed \sys to offload the signing of evidence to a dedicated trusted kernel module, called the attestation service (0.5\,k SLOC).
It plays a critical role in \sys as it has access to the private attestation key.
The attestation service, located in the kernel space of OP-TEE, prevents the key materials from being exposed to the TAs in the user space.
Hence, the \wasm runtime communicates claims to the attestation service for evidence generation.

In order to establish a root of trust, our hardware is equipped with a cryptographic accelerator and assurance module (CAAM).
The root of trust is a unique 256-bit one-time programmable key (OTPMK), fused into hardware at manufacturing time.
The CAAM provides two different hashes of OTPMK, depending on if the requesting thread is in the normal or in the secure world. This hash is called the master key verification blob (MKVB).
The MKVB is then used as a seed to provision secrets only known by a kernel module in OP-TEE~\cite{imx8securityreference, opteereadforproduction}.
We modified OP-TEE to expose the total size of the MKVB as it only supported 128-bit hardware keys.

\ifnum\isExtendedPaperEnabled=1
\fi

We used the library LibTomCrypt~\cite{libtomcrypt} for cryptographic operations, since OP-TEE already uses it.
We decided to use elliptic-curve cryptography (ECC) to reduce the key size for faster transmission and lower processor consumption, while offering the same level of security compared to RSA with a large modulus~\cite{1269719}.
We selected the curve \emph{secp256r1} as recommended by the NIST~\cite{Nist2020Recommendation}, as well as the following algorithms:
\begin{enumerate*}[label=\emph{(\roman*)}]
    \item elliptic-curve digital signature algorithm (ECDSA) (256-bit) for the attestation key pair,
    \item ephemeral elliptic-curve Diffie–Hellman (ECDHE) (256-bit) for the session keys of the remote attestation protocol,
    \item AES-GCM (128-bit) for data encryption and authentication,
    \item SHA (256-bit) for the anchor in the evidence, and
    \item AES-CMAC (128-bit) for the message authentication code (MAC) of the messages, to derive the KDK and shared keys.
\end{enumerate*}

We also extended OP-TEE to generate an attestation key pair at each boot, in a deterministic manner, based on the hardware root of trust. To accomplish this task, we modified LibTomCrypt in OP-TEE to include a pseudorandom number generator (PRNG) named \emph{Fortuna}, as the OP-TEE's PRNG does not support seeds.
Furthermore, we changed the wrapper of LibTomCrypt to generate an ECC key pair with a seed fed into Fortuna.
The current implementation of the wrapper for key generations relies on a single instance of PRNG that uses hardware randomness.
The generation of the ECDSA key pair is done in two steps:
first the MKVB is derived using the built-in function \texttt{huk\_subkey\_derive},
then the resulting value is used as a seed for Fortuna, and the ECDSA key pair is generated using that PRNG instance.

\smallskip\noindent{\textbf{Extension to WASI: WASI-RA.}}\quad
We propose WASI-RA, an extension to WASI for remote attestation, implementing the mechanism designed in \sys.
This enables hosted \wasm applications to control the remote attestation flow.
In the remainder of this section, we briefly describe these functions. 

The \wasm runtime exposes two functions for evidence generation to \wasm applications: \texttt{wasi\_ra\_collect\_quote} and \texttt{wasi\_ra\_dispose\_quote}.
The former issues evidence based on an anchor given as a parameter, ensuring freshness and uniqueness.
The evidence is returned in the form of an opaque handle.
The latter requires a handle of evidence and disposes of it.
These two functions are deliberately not coupled to the attestation protocol to be used with other transport layers (\eg TLS).

The remaining WASI-RA functions implement the attestation protocol.
\texttt{wasi\_ra\_net\_handshake} sends and receives $\text{msg}_0$ and $\text{msg}_1$ (\ie the handshake), according to a host address and the identity (public ECDSA key) of the verifier.
The remote party's identity is usually hardcoded in the application, so the code measurement enables the server to detect whether it has been maliciously changed.
A remote attestation context and an anchor are returned in opaque values; the latter is used to generate evidence.
The functions \texttt{wasi\_ra\_net\_send\_quote} and \texttt{wasi\_ra\_net\_receive\_data} send the evidence ($\text{msg}_2$) and receive the secret blob ($\text{msg}_3$), respectively.
The secret blob is retrieved as a byte array with a variable size.
Finally, the context of the remote attestation is disposed of using \texttt{wasi\_ra\_net\_dispose}. 

\section{Evaluation}
\label{sec:evaluation}

Our evaluation answers the following questions:
\begin{enumerate*}[label=\emph{(\roman*)}]
    \item are time measurements sufficiently accurate inside TrustZone, and if not, how can we improve this?
    \item what is the performance overhead for compute-bound \wasm applications in TrustZone?
    \item how do real-world applications, compiled as \wasm, perform in TrustZone?
    \item what are the cost factors of our RA protocol?
    \item what is the overhead of running a machine learning application using the RA protocol?
\end{enumerate*}
We respectively answer these questions by assessing the usage of monotonic timers (\S\ref{sec:eval-measure-time}), using a general-purpose computing-bound evaluation with \polybench (\S\ref{sec:micro-wasm}), evaluating SQLite as a real-world embeddable database (\S\ref{sec:macro-wasm}), breaking down the cost of our remote attestation operations (\S\ref{sec:micro-ra}) and assessing an end-to-end machine learning scenario using Genann (\S\ref{sec:macro-ra}).

\emph{Experimental setup.}
All experiments run on an off-the-shelf NXP MCIMX8M evaluation board, equipped with i.MX 8MQ, an Arm Cortex-A53 (1.5\,GHz) SoC with the Armv8-A architecture.
In the \$100 price range, this board supports the hardware root of trust, secure boot, and the CPU fully supports TrustZone.
The normal world OS is compiled using BuildRoot 2021.02, with the Linux kernel 5.13 forked by Linaro to ensure compatibility with OP-TEE.
The secure world OS runs OP-TEE 3.13.
The bootloaders are U-Boot 2020.10-rc2 and Arm Trusted Firmware 2.3.

We show the median and standard deviation of multiple runs for each experiment, as specified with each benchmark. 
For most experiments, the standard deviation is very small.
The native benchmarks are compiled, using GCC 9.2.1 and \texttt{-O3} optimisation.
The \wasm benchmarks are compiled by WASI-SDK, which uses Clang 11 with the same optimisation flag.
Our implementation is open-source, and instructions to reproduce our experiments are available on GitHub~\cite{watz-repo}.

\subsection{Time measurements in TrustZone}
\label{sec:eval-measure-time}

\begin{figure}[!t]
    \includegraphics[width=\columnwidth]{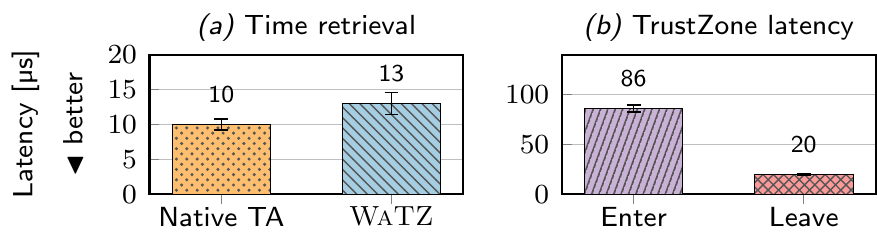}
    \caption{Time retrieval and world transition latencies.}
    \label{fig:eval-measure-time:latencies}
    \vspace{-12pt}
\end{figure}

The time resolution of the measurements inside TrustZone, offered by default from OP-TEE, is in milliseconds.
To achieve nanosecond resolution, we implemented and evaluated the following changes across all involved software stacks to retrieve the same time as provided by the monotonic clock of the Linux kernel in the normal world.
For native applications compiled for the normal world, the time is returned using the standard POSIX function \texttt{clock\_gettime}.
However, native applications in the secure world do not have a standardised way to retrieve the same time.
Hence, we modified the OP-TEE driver to add a function for passing the value of the monotonic clock to the secure world.
We also extended the GP's type \texttt{TEE\_Time} to measure our experiments with a nanosecond precision.
Finally, \wasm applications rely on WASI with the function \texttt{clock\_time\_get}.
Doing so, WASI-SDK~\cite{wasisdk} seamlessly maps it to the POSIX \texttt{clock\_gettime}.

Trusted applications require to specify the heap and stack sizes at compile time.
As such, we allocated a heap size of 2\,MB and a stack size of 3\,KB for this benchmark.
This amount comprises the memory of the runtime, the bytecode of the \wasm application and the space of the virtual heap and stack allocated by \sys to execute the \wasm program.

\autoref{fig:eval-measure-time:latencies}\hyperref[fig:eval-measure-time:latencies]{a} shows the latencies to fetch the time in two settings, respectively, from native trusted applications and \wasm in TEE.
We ran each experiment 1000 times.
The time required in normal world for native and \wasm programs is under 1\microsec (not shown).
The average latency of retrieving the time in the secure world is \bmLatenciesTimeNativeTaTimeMean\microsec for native applications and \bmLatenciesTimeWasmTaTimeMean\microsec for \wasm applications.
The increase is due to a transition to the normal world for each query.
The benchmarks presented in the following of this section take such latency into account.

\autoref{fig:eval-measure-time:latencies}\hyperref[fig:eval-measure-time:latencies]{b} shows the time to switch between worlds, a frequent operation when an application is partitioned to execute sensitive operations in TrustZone.
In \sys, the server of the verifier invokes functions inside the TEE once received by the TCP server.
Our micro-benchmark registers the time in the normal world, before and after the TEE invocation.
Similarly, we measured the time in the secure world, upon a function call.
We observed an average time of \bmLatenciesRoundtripEnterMean\microsec to call a function in the secure world, and \bmLatenciesRoundtripLeaveMean\microsec to return, as observed earlier~\cite{10.1007/978-3-030-22496-7_9}.

\subsection{Startup Overhead}
\label{sec:startup:overhead}

\begin{figure}[t]
    \includegraphics{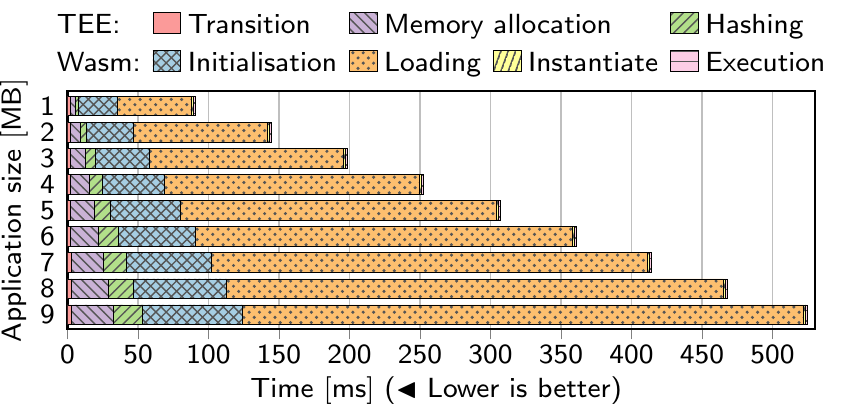}
    \caption{Startup breakdown of \wasm applications in \sys.}
    \label{fig:eval-measure-time:launch-time}
    \vspace{-12pt}
\end{figure}

\begin{figure*}
    \centering
    \includegraphics{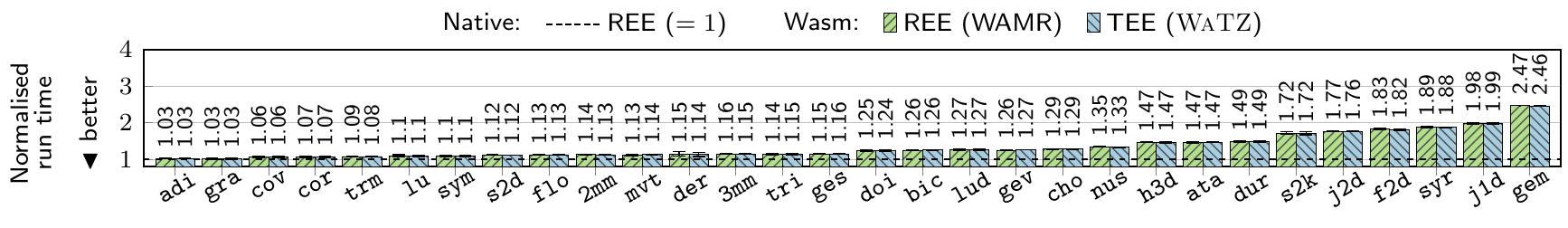}
    \caption{Performance of \polybench, normalised against native (unsecure/REE world).}
    \label{fig:polybench}
\end{figure*}

\begin{figure*}
    \centering
    \includegraphics{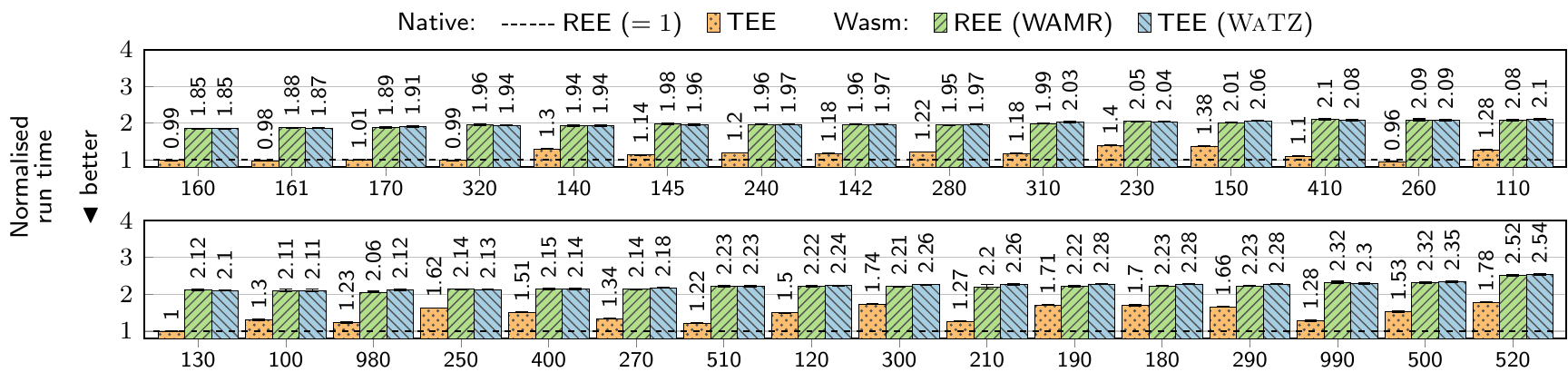}
    \caption{Performance of SQLite's Speedtest1, normalised against native (unsecure/REE world).}
    \label{fig:speedtest1}
\end{figure*}

Next, we evaluated the startup overhead of \wasm applications loaded into \sys.
For that purpose, we created nine \wasm programs with a code size that varies from 1 to 9\,MB.
The AOT compiled \wasm binaries have been generated by unrolling thousands of loop iterations to output an amount of code with a size of 1\,MB.
That output is replicated multiple times to create the nine variants.
Afterwards, we measured the time from the instruction that requests the launch of \wasm application in the normal world until the first instruction in the \wasm application is executed.
Finally, the \wasm program stops after the first \wasm instruction since we measure the startup time, preventing the loops from being executed.

We allocated a heap size of 23\,MB, which is the smallest amount of memory to launch the application of the largest size (9\,MB).
We identified that the overhead of memory, roughly twice the application code size, was bound to the relocation symbols LLVM generated in the AOT compiled \wasm programs.
The \wasm runtime (WAMR) allocates a dedicated structure for each relocation entry.
Hence, we determined that the loading operation of an AOT compiled \wasm binary in the runtime may double the size to allocate, depending on the structure of the code.
One way to reduce that overhead is the use of the experimental feature of the ahead-of-time compiler called \emph{execution in place}, which generates as few as possible AOT relocations.
We leave this optimisation as future work since this feature still has known issues.

\autoref{fig:eval-measure-time:launch-time} shows a breakdown summary of the various internal operations when loading the previously mentioned \wasm applications within \sys.
The large majority is dedicated to allocating the memory for the \wasm AOT bytecode (\bmLaunchTimeMemoryRatio\%), initialising the \wasm runtime (\bmLaunchTimeWasmInitRatio\%) and loading the bytecode (\bmLaunchTimeWasmLoadRatio\%).
The memory allocation creates two buffers: \emph{(1)} to store the \wasm AOT bytecode in the secure memory and, \emph{(2)} for the heap of the \wasm application.
The initialisation of the runtime consists of creating the \wasm runtime environment, initialising the memory allocator, and registering native symbols (\ie the binding of the native functions imported by the \wasm application). 
The loading phase parses the bytecode and creates the internal structures required to run \wasm applications, similarly to loading a normal world process.
This phase notably includes the loading of the relocation entries.
Hashing the bytecode takes \bmLaunchTimeHashRatio\% of the time on average.
The hash is later embedded in the evidence issued during the remote attestation process.
Each of the remaining categories (\ie, the time to transition to the secure world, \wasm instantiation and execution) takes less than 1\% of the startup time.
Compared to the baseline \wasm runtime in the normal world (WAMR), the overhead added by \sys is the transition time and the hashing operation, which represents roughly an increase of 5\%.

\subsection{\wasm micro-benchmarks: \polybench}
\label{sec:micro-wasm}

\polybench~\cite{polybench} is a CPU-bound benchmark suite commonly used to evaluate compiler optimisations~\cite{10.1007/978-3-662-45231-8_41} and often used for comparative analysis of \wasm environments~\cite{10.1145/3361525.3361541,234914,254432}.
We use all the tests in the \polybench suite (v4.2.1b), executed individually 50 times.
We compare \wasm executed ahead-of-time in the normal world (using WAMR) and in the secure world (using \sys), against native execution (Arm, baseline).
Due to memory limitations in the secure world imposed by OP-TEE, we rely on the built-in medium dataset of \polybench for all 30 applications.
We allocated a heap size of 12\,MB for \sys, which is sufficient to execute all the tests of the benchmark suite.

\autoref{fig:polybench} shows these results.  
In both normal and secure worlds, we observe that \wasm's slowdown is on average  \bmPolybenchTeeGlobalSlowdown$\times$ when compared to native execution.
Therefore, \sys does not add additional penalties when executed in the secure world.
The difference observed between WAMR and \sys are insignificant (less than 0.02\%), since TrustZone does not introduce any security mechanism that slows down the computation speed.
These results confirm previous work comparing the slowdown of \wasm applications against their native counterpart using JIT and AOT compilation~\cite{twine,234914}.    
A qualitative comparison of our results with those found for \polybench in Intel SGX~\cite{twine} shows that TrustZone does not affect the execution runtime negatively.
Unlike our proposal, Intel SGX introduces noticeable performance overheads on AOT compiled \wasm execution.
We think the difference is natural because of the additional security guarantees provided by SGX, with transparent encryption and verification of memory pages stored in volatile memory~\cite{cryptoeprint:2016:086}.

\subsection{\wasm macro-benchmarks: SQLite}\label{sec:macro-wasm}

SQLite~\cite{5231398} is a widespread, portable, low memory footprint, full-fledged embeddable database, well suited for constrained environments.
Our tests use SQLite v3.36, leveraging its built-in benchmarking suite Speedtest1~\cite{speedtest1}.
We use the native execution of SQLite (\armarch) in the normal and secure worlds as an empirical baseline.
As such, we adapted SQLite to be embedded within a trusted application and run in OP-TEE.
We instantiate exclusively in-memory databases, as we have left the implementation of WASI's file system API for future work.
Each benchmark in the suite assesses a single aspect of the database engine (\eg selections with joins, updates of data or changes of the schema, \etc).
We configured SQLite to use a 2048-page cache of 4\,kB each (for a cache size of 8~MB), with the default (normal) synchronous mode and the default (delete) journal mode.
In addition, we enabled an alternate memory allocator to pre-allocate the memory used by the database.
To fit in the restricted memory of the secure world imposed by OP-TEE, we scale down the input dataset to 60\% (argument \texttt{--size}).
As such, we compiled the trusted application of \sys to use 25\,MB of heap memory.

Notice that TrustZone does not impose such memory limitations.
Hence the entire dataset can be used as soon as OP-TEE lifts memory restrictions.
Provided the memory can be enlarged, we do not foresee any impediment for \sys to operate on software with large memory footprints, such as deep learning systems for instance.
The experiments ran 50 times, and we report medians and standard deviations in \autoref{fig:speedtest1}.

Results are normalised against the execution time of Speedtest1 in the normal world, using native execution as a baseline.
Overall, our observations are twofold: \emph{(1)} in the normal world, the slowdown of WAMR on average is \bmSpeedtestOneGlobalSlowdownReeWasm$\times$, and \emph{(2)} in the secure world, the slowdown of native execution and \sys on average are \bmSpeedtestOneGlobalSlowdownTeeNative$\times$ and \bmSpeedtestOneGlobalSlowdownTeeWasm$\times$ respectively.
As a result, the execution speed overhead of \wasm over native execution in the TEE on average is \bmSpeedtestOneGlobalRatioTeeWasmAndTeeNative$\times$.
This macro-benchmark also demonstrates that \sys has low to no-overhead execution in TrustZone, compared to \wasm in the normal world.
Conversely, the native TA suffers from an overhead compared to the native application in the normal world.
These slowdown differences are explained because the compiled binary in the normal world is optimised for using the underlying hardware, unlike \wasm applications that are ahead-of-time compiled from the intermediate \wasm bytecode.
Most of the experiments located on the top of \autoref{fig:speedtest1} perform read queries on the database and have a lower impact on the performance, with an average of \bmSpeedtestOneSelectAverage$\times$ (\ie experiments 130-145, 160-170, 260, 310, 320, 410, 510, 520).
Instead, most of the bottom experiments are write-intensive, by inserting data in the database structure, with a slowdown on average of \bmSpeedtestOneInsertAverage$\times$ (\ie experiments 100-120, 180, 190, 210, 290, 300, 400, 500).
Similarly to the micro-benchmarks, we identify minor discrepancies where the TEE applications are faster (equal to or less than 0.04\%).
We also consider them insignificant and are deemed equal to the performance measured in their unsecured counterpart.

In summary, our analysis of \wasm runtime performance in \sys shows overheads of \bmPolybenchTeeGlobalSlowdown$\times$ and \bmSpeedtestOneGlobalSlowdownTeeWasm$\times$ for the micro- and the macro-benchmarks, respectively.
We estimate that the security benefits of TrustZone largely compensate for these performance penalties.
Besides, we did not observe any noticeable overhead using \sys compared to WAMR.
Thanks to WebAssembly, our tool provides great freedom for the developers who build secure software using various technologies without being bound to the programming restrictions of OP-TEE.
On this topic, we noted that the adaptation of SQLite to be executed in OP-TEE as a native TA has been substantially more laborious than compiling it in \wasm with WASI support.
Indeed, the specialised API offered for trusted applications required to modify the source code of SQLite at various locations to replace missing POSIX functions and undefined data structures.
On the other hand, WASI-SDK, the framework used to compile C programs in \wasm, provides all the function and structure definitions needed for a seamless compilation.
Finally, we leveraged the robust sandbox of \sys to ensure isolation between each hosted application in the secure world with acceptable costs.

\def\labelcolumnlength{35mm}
\newcommand{\generation}{\textsuperscript{\textdagger}}
\newcommand{\handling}{\textsuperscript{\textdaggerdbl}}
\begin{table}[t]
    \captionsetup{type=table}
    \small
\setlength{\tabcolsep}{2pt}
    \rowcolors{1}{gray!10}{gray!0}
    \begin{tabularx}{\columnwidth}{p{\labelcolumnlength}RRR}
    \toprule
    \rowcolor{gray!25}
    \textit{(a)~Attester} &\generation $\text{msg}_0$                               &\handling $\text{msg}_1$                                     &\generation $\text{msg}_2$\\
    \midrule
    Memory management                   & \bmAttesterMemMessageZeroMeanInMicrosec\microsec                 & \bmAttesterMemMessageOneMeanInMicrosec\microsec                      & \bmAttesterMemMessageTwoMeanInMicrosec\microsec\\
    Key generation                      & \Ding{1} \bmAttesterKeygenMessageZeroMeanInMs\ms     & \Ding{5} \bmAttesterKeygenMessageOneMeanInMs\ms          & --- \\
    Symmetric cryptography              & ---                                           & \bmAttesterSymCryptoMessageOneMeanInMicrosec\microsec                & \bmAttesterSymCryptoMessageTwoMeanInMicrosec\microsec\\
    Asymmetric cryptography             & ---                                           & \Ding{4} \bmAttesterAsymCryptoMessageOneMeanInMs\ms      & \Ding{6} \bmAttesterAsymCryptoMessageTwoMeanInMs\ms\\
    \bottomrule
    \rowcolor{gray!25}
    \textit{(b)~Verifier} &\handling $\text{msg}_0$                                 &\generation $\text{msg}_1$                                   &\handling $\text{msg}_2$\\
    \midrule
    Memory management                   & \bmVerifierMemMessageZeroMeanInMicrosec\microsec                 & \bmVerifierMemMessageOneMeanInMicrosec\microsec                      & \bmVerifierMemMessageTwoMeanInMicrosec\microsec\\
    Key generation                      & \Ding{2} \bmVerifierKeygenMessageZeroMeanInMs\ms     & ---                                               & ---\\
    Symmetric cryptography              & ---                                           & \bmVerifierSymCryptoMessageOneMeanInMicrosec\microsec                & \bmVerifierSymCryptoMessageTwoMeanInMicrosec\microsec\\
    Asymmetric cryptography             & ---                                           & \Ding{3} \bmVerifierAsymCryptoMessageOneMeanInMs\ms      & \Ding{7} \bmVerifierAsymCryptoMessageTwoMeanInMs\ms\\
    \bottomrule
    \multicolumn{4}{l}{\cellcolor{gray!0}\generation\scriptsize{Generation of the message.}\quad\cellcolor{gray!0}\handling\scriptsize{Handling the message.}}\\
    \end{tabularx}
    \rowcolors{1}{gray!10}{gray!0}
    \vspace{5pt}
    \caption{Execution time of $\text{msg}_{0}$, $\text{msg}_{1}$ and $\text{msg}_{2}$.}
    \label{tbl:ra-micro:messages012}
    \vspace{-25pt}
\end{table}

\subsection{Remote attestation micro-benchmarks}
\label{sec:micro-ra}

We now evaluate the remote attestation protocol.
For the sake of these tests, the attester (client) and the verifier (server) applications run on the same development board. 

We measure the execution time of each message in the remote attestation protocol and highlight the various cost factors.
The sizes of $\text{msg}_0$, $\text{msg}_1$ and $\text{msg}_2$ are fixed, and the execution time of their generation and handling are not bound to the size of the Wasm application being measured.
On the other hand, $\text{msg}_3$ requires a time that is proportional to the size of the secret blob transferred to the application.
As a consequence, we analyse the execution time of the three first messages regardless of Wasm AOT bytecode size in \autoref{tbl:ra-micro:messages012}.
\autoref{fig:ra-micro:message3} depicts the execution time of the fourth message, according to the size of data to transmit securely.
We evaluate the transfer of confidential information, from 512\,kB to 3\,MB.
We compiled the attester and the verifier of \sys to approximately split the heap memory equally as they are located on the same hardware: 14\,MB and 13\,MB, respectively.
We noticed that our implementation requires twice the size of the transmitted data in memory, in each trusted application.
Indeed, we needed to allocate a buffer for the plaintext and a buffer for the ciphertext.
For the largest data to transfer (3\,MB), this represents 6\,MB in each TA, leading to a memory occupancy of 12\,MB in total.
We leave the optimisation of encryption and decryption using a single buffer for future work.
In addition to the details below, we report that most of the time is dedicated to complete asymmetric cryptography operations, \ie keys and signatures generation.

\paragraph{Message 0}
The attester generates an ECDHE key pair and sends the public key (\Ding{1} in \autoref{tbl:ra-micro:messages012}).
When handling message 0, the verifier generates an ECDHE key pair and derives the shared secrets to establish a secure channel (\Ding{2}). 

\paragraph{Message 1}
The verifier signs the ECDHE public keys (\Ding{3}) and generates a MAC.
The asymmetric signature takes most of the time, up to $\numprint{\bmMessageOneAsymVsSymRatio}\times$ the execution time of the (symmetric) MAC, as expected~\cite{paar2009understanding}.
The derivation of the shared secrets is also performed on the attester's side when handling $\text{msg}_1$ (\Ding{5}).
Hence the time the attester takes to generate its keys (\Ding{1}, \Ding{5}) is similar to the time the verifier takes to do the same (\Ding{2}).

\begin{figure}[t]
    \vspace{-5pt}
    \includegraphics{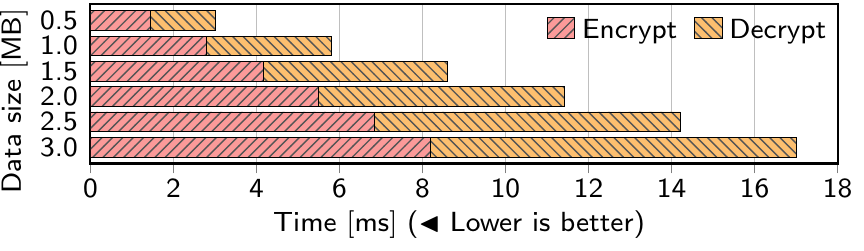}
    \vspace{-10pt}
    \caption{Execution time of $\text{msg}_3$.}
    \label{fig:ra-micro:message3}
    \vspace{-10pt}
\end{figure}

\paragraph{Message 2}
The attester issues the evidence, which requires a digital signature (\Ding{6}), and a MAC.
Upon reception, the verifier checks the MAC and the evidence signature (\Ding{7}).
The time to sign (\Ding{6}) and verify the signature (\Ding{7}) of $\text{msg}_2$ is similar to the time required for the same operations on $\text{msg}_1$ (\Ding{3} and \Ding{4}).
The same cryptographic operations are performed in both cases, using different data.

\paragraph{Message 3}
Finally, the verifier generates $\text{msg}_3$ with the secret blob, encrypted and authenticated using AES-GCM.
We omitted the time consumed by memory management (less than 1\%).
As seen in \autoref{fig:ra-micro:message3}, the execution time evolves proportionally between the verifier's encryption and the attester's decryption.
Since AES-GCM is symmetric, it is faster than signing the three first messages, starting from \bmMessageThreeMinimumTime\ms for 0.5\,MB of encrypted data and reaching \bmMessageThreeMaximumTime\ms for 3\,MB.

\subsection{Remote attestation macro-benchmarks: Genann}\label{sec:macro-ra}

\begin{table}[!t]
    \small
\setlength{\tabcolsep}{2pt}
        \begin{tabularx}{\columnwidth}{RP{70pt}P{65pt}P{40pt}}
        \toprule
        \rowcolor{gray!25}
        \texttt{handshake}                             & \texttt{collect\_quote}                   & \texttt{send\_quote}                                      & {\footnotesize Baseline}\\
        \midrule
        \bmGenannNetHandshakeMean\s                                          & \bmGenannCollectQuoteMean\ms                 & \bmGenannSendQuoteMean\ms                                    & \cellcolor{gray!10} \bmGenannSumOfWasiRaWithoutData\s\\
        \bottomrule
        \rowcolor{gray!25}
        {\footnotesize  $\hookrightarrow$ Baseline}   & {\footnotesize Size}                      & \texttt{receive\_data}                                    & {\footnotesize Total}\\
        \midrule
        \rowcolor{gray!10}
        \bmGenannSumOfWasiRaWithoutData\s      &\cellcolor{gray!0}{\footnotesize 0.1~MB}   & \cellcolor{gray!0}\bmGenannReceiveDataMinMeanInMs\ms   & \bmGenannSumOfWasiRaMin\s\\
        \rowcolor{gray!10}
        \bmGenannSumOfWasiRaWithoutData\s      &\cellcolor{gray!0}{\footnotesize 1~MB}     & \cellcolor{gray!0}\bmGenannReceiveDataMaxMeanInMs\ms   & \bmGenannSumOfWasiRaMax\s\\
        \bottomrule
        \end{tabularx}
        \vspace{6pt}
        \caption{Execution time of WASI-RA API.}
        \label{tbl:genann}
        \vspace{-10pt}
\end{table}

We conclude our evaluation of \sys with Genann~\cite{genann}, a neural network library extensively used in literature~\cite{cantoro2020evaluating,amornpaisannon2020laser}. This library supports feedforward artificial neural networks (ANN) and has zero external dependencies, making it a convenient target to be compiled in \wasm and tested in a constrained memory environment.
The benchmark is based on a built-in Genann example, where an ANN is trained on a subset of the Iris dataset~\cite{Dua:2019}.
The ANN comprises 4 inputs, 1 hidden layer of 4 neurons and 3 outputs (1 per class).
The input dataset includes 50 records per class (file size of 4.45\,kB).
We replicated the dataset to reach the breakpoint sizes, from 100\,kB to 1\,MB. 
Tests are executed 20 times.
The attester is launched with Genann as a trusted \wasm application.
We allocated 17\,MB for the attester, enough space to handle the \wasm runtime, the attestation with the transfer of the dataset and the heap required by Genann.
The remaining memory is allocated for the verifier (10\,MB).
Once executed, it triggers a remote attestation request to retrieve the dataset, used to train and predict the classification.
We use this end-to-end example to demonstrate a real-world workflow using \sys and assess a few cost factors, such as the impact on the execution time when the size of the confidential information varies.
Similarly to the micro-benchmarks of remote attestation (\S\ref{sec:micro-ra}), the attester and verifier are co-located on the same device.

\autoref{tbl:genann} shows the execution time of WASI-RA, the API exposed to the \wasm applications to request remote attestation and evidence generation.
The function retrieving the secret blob is indicated according to the lower and upper bounds of the dataset size.
This end-to-end benchmark includes client and server time to generate and handle the messages, the overhead caused by the socket connection and the penalty of the normal and secure world switching.

Most of the execution time is spent on the handshake: $\text{msg}_0$ and $\text{msg}_1$ handle the key generation and half on the asymmetric operations, as seen in \autoref{tbl:ra-micro:messages012}.
The generation of the evidence is the second most time-consuming operation due to the digital signature (\Ding{7} in \autoref{tbl:ra-micro:messages012}).
The sending of the evidence consumes only a marginal time.
Lastly, we evaluated several dataset sizes to assess the execution time of the function that receives the confidential information.
We report an execution time ranging from \bmGenannReceiveDataMinMeanInMs\ms for 100\,KB up to \bmGenannReceiveDataMaxMeanInMs\ms for 1\,MB.
According to the micro-benchmark in \autoref{sec:micro-ra}, the cryptographic operations of $\text{msg}_3$ takes a negligible amount of time: \bmMessageThreeTimeAtOneMB\ms for 1\,MB of data, unlike this macro-benchmark.
This difference is due to the attester waiting for the end of the evidence verification on the verifier's side, reported for lasting \bmVerifierAsymCryptoMessageTwoMeanInMs\ms in \autoref{tbl:ra-micro:messages012} (\Ding{7}).
We confirmed this was the cause by waiting before receiving the data, which leads to a reception duration of only 70\ms for 1\,MB.
This points out the importance of having appropriate hardware for cryptographic operations on the verifier's side.
Further, we evaluate the training time of the model (\autoref{fig:genann:training}) for different dataset sizes.
The dataset is fetched from a regular file in the normal world (using WAMR).
In \sys, it is read via a secure channel established during the remote attestation.
\sys achieves a \bmGenannTrainingSpeedUpPercent\% speedup, roughly matching previous experiments with no runtime overheads.
The time difference between the two systems is yet to be further investigated.
Indeed, using the same AOT binary and supplying the same data to Genann's training function yields better performance in OP-TEE.
We excluded the cause of better benchmark scheduling, since OP-TEE relies on the REE Linux scheduler.
Finally, we report how the performance obtained here confirms similar results in literature~\cite{10.1145/3098954.3098971}, where an equally powerful TrustZone hardware (HiKey Arm Cortex-A53, at 1.2\,GHz) is used to benchmark remote attestation protocols using TrustZone.

\begin{figure}[!t]
    \centering
    \vspace{-9pt}
    \includegraphics[width=\columnwidth]{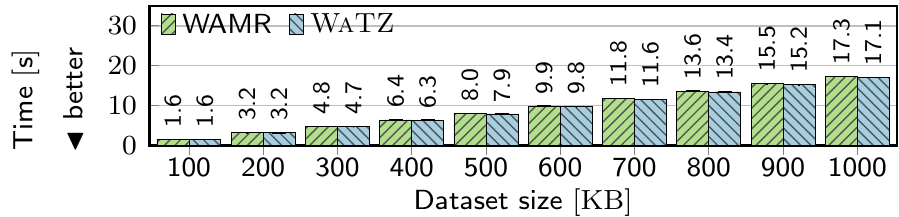}
    \caption{Execution time of the training phase.}
    \label{fig:genann:training}
    \vspace{-11pt}
\end{figure}

\section{Security analysis}
\label{sec:security}

This section provides a security analysis of the concepts within \sys and discusses their strengths and limitations.

\smallskip\noindent{\textbf{Secure boot.}}\quad
The ROM and the eFuses ensure that the board is booting the secure OS in a genuine state.
Assuming an attacker replaces the image of the trusted OS, the software signature will no longer match the binaries, which aborts the boot sequence.
Since the signature's public key is stored in the eFuses, it cannot be altered to install a compromised system.
Consequently, only a genuine version of the secure kernel OS can access the root of trust.
Nonetheless, the system does not protect against rollback attacks.
This can be locally mitigated using monotonic counters bound to the hardware, and remotely mitigated by checking the version of \sys in the evidence during remote attestation requests.
While not available in the SoC of our evaluation board, the security guarantees of \sys may be further extended using a measured boot in addition to the secure boot.
Measured boot is a security mechanism that enables the attester to collect the code measurement of every loaded software (\eg bootloader, trusted OS) and accumulate them in special registers.
Once these values (considered additional claims) are collected, they are signed and exported, typically using a trusted platform module (TPM).
These claims may be later incorporated into evidence produced by \sys as a system-wide measurement.
Consequently, verifiers may also appraise the startup components to identify a potentially hijacked secure boot.

\smallskip\noindent{\textbf{Trusted \wasm runtime.}}\quad
The runtime offers execution and attestation facilities similar to Intel SGX.
The normal world requests \sys to start \wasm applications and, as a result, they are executed in isolation inside TrustZone.
Due to the memory sandboxing inherited from \wasm, malicious applications loaded in \sys cannot compromise the secure world or access resources owned by other applications.
This aspect is different from the regular TAs in OP-TEE, because a signing key is needed to deploy software in the secure world, which prevents the execution of arbitrary software~\cite{opteetasigning}.
Moreover, the communication of the signing key to allow third-party developers to execute trustworthy code in OP-TEE may lead to impersonation attacks on already deployed applications, ultimately leaking secrets stored in persistent storage~\cite{opteetaimpersonation}.
As such, the \wasm sandbox offers a safe and efficient isolation mechanism to host software made by different parties.
The reliance on a runtime has the downside of increasing the size of the trusted computing base.
Furthermore, zero-day vulnerabilities found in the \wasm sandbox may lead to data leakage or arbitrary code execution, because the isolation offered by \wasm is implemented in software, as opposed to the isolation mechanisms of SGX enclaves.
We note this risk also applies to the trusted applications of OP-TEE, since vulnerabilities found in the trusted OS may lead to privilege escalations and, therefore, weaken the isolation of the trusted applications.
We mitigate this issue by providing the version of \sys in the evidence, so the verifier can determine whether the deployed runtime is up to date, \ie detect whether the attester has mitigation patches applied.

\smallskip\noindent{\textbf{Remote attestation protocol.}}\quad
\label{sec:security:protocol}
We evaluated the correctness of our remote attestation protocol with Scyther~\cite{10.1007/978-3-540-70545-1_38}, a state-of-the-art formal analyser.
This tool is known to analyse security protocols, security requirements and identify vulnerabilities formally.
Scyther is based on the protocol semantics model for the Dolev-Yao intruder model~\cite{1056650}, which assumes that an adversary has complete control over the communication channel (an attacker can do everything except breaking cryptography).
We configured Scyther to check the secrecy of the private session keys, the shared secret and the secret blob.
Besides, we verified the following claims of authentication: \emph{aliveness}, \emph{weak agreement}, \emph{non-injective agreement}, \emph{non-injective synchronisation} and \emph{reachability} (\ie the protocol ended on both parties).
While we omit to present these terms for conciseness~\cite{20.500.11850/58054,596782}, they represent essential characteristics of a security protocol.
Scyther revealed no attack or flaw in our proposal.
The script of the protocol is available in the repository of \sys~\cite{watz-repo}.

\section{Conclusion}
\label{sec:conclusion}

To the best of our knowledge, \sys is the first \wasm runtime executing entirely inside Arm TrustZone with full support for remote attestation, optimised explicitly for \wasm to establish trust on hosted applications.

In many ways, \sys offers a model similar to Intel SGX while overcoming the limitations of TrustZone, extending the paradigm of a single trusted world into fully isolated and mutually distrusting enclaves.
It comprises a trusted environment for running unmodified \wasm applications that cannot be tampered with.
It relies on a hardware root of trust and a secure boot, shielded by Arm TrustZone.
Our WASI extensions for remote attestation and the attestation protocol establish trust on remotely executed \wasm applications inside \sys and securely provision confidential data, such as shared keys to join a channel or decrypt a configuration file.
Our extensive experimental evaluation assesses the costs of the \sys mechanisms with typical tasks IoT devices carry out.
As a result, our implementation is lightweight, achieving results on par with \wasm running in the normal world.

We have a strong commitment to open-source and reproducibility: some of our improvements are already pushed to the open software projects we depend on, and we intend to keep doing so.
The code of \sys and the experiments presented in this paper are available on GitHub~\cite{watz-repo}.

\ifnum\isExtendedPaperEnabled=1
\fi

\section*{Acknowledgments}
This publication incorporates results from the VEDLIoT project, which received funding from the European Union’s Horizon 2020 research and innovation programme under grant agreement No 957197.

\bibliographystyle{IEEEtran}

\end{document}